\documentclass[10pt]{article}

\usepackage[utf8]{inputenc}
\usepackage{geometry}
\usepackage{setspace}
\usepackage{titlesec}
\usepackage{lmodern}
\usepackage[numbers]{natbib}
\usepackage{graphicx}
\usepackage{hyperref}
\usepackage{csquotes}
\usepackage{multicol}
\usepackage{parskip}
\usepackage{enumitem}
\usepackage{subcaption}

\titleformat{\section}[block]{\Large\bfseries}{\thesection.}{0.5em}{}
\titleformat{\subsection}[block]{\large\bfseries}{\thesubsection}{0.5em}{}

\title{\textbf{Seeing Beyond Sound: \\Visualization and Abstraction \\in Audio Data Representation}}
\author{Ashlae Blum’e \\ \texttt{ashlaeblume@gmail.com}}
\date{July 2025}

\begin{document}
\maketitle


\vspace{1em}

\begin{abstract}
The interpretation of complex data is epistemically linked to human perceptual frameworks. In audio information research, sound is represented and transformed using visual elements that highlight abstract patterns detached from the physical experience of perception. While ubiquitous throughout audio research domains, software tools carry hidden assumptions that are inherited from their historical contexts. However, these conventions are often masked through their adoption to new scientific uses. For audio data, waveforms, spectrograms, and DAW-like (digital audio workstation) interfaces are the cornerstones of interactive visualization. However, the visual presentation of information strongly influences an individual’s ability to form complex associations. As such, modern audio data workflow requirements run the risk of misalignment with tools that were originally designed for other uses. We argue that re/designing tools to align with emergent needs of modern users will improve both analytical as well as creative outputs due to an increased affinity for using them. This paper explores the potentials associated with adding dimensionality back into visualizations to facilitate positive outcomes in the use of audio information visualization tools. 
\end{abstract}

\newpage

{\small
\begin{center}
\textbf{Inscription on the Wall of West Forest Temple}

Viewed from the front, a full mountain range; from the side, a single peak. \\
Far, near, high, low – each view is different. \\
I cannot recognize the true face of Mount Lu, \\
Simply because I myself am on the mountain. \\
– Su Shi
\end{center}
}

\section{Introduction}
Advanced data visualization techniques let scientists interpret complex datasets by transforming high-dimensional data into abstract visual elements. This reveals patterns in information and builds narratives that enhance collective understanding. For audio data, waveforms and spectrograms form the basis of our visual knowledge. These rely on two-dimensional depictions of the time-frequency domain that are mathematically well-defined, but often lack intuitive correspondence with the multisensory nature of auditory perception. The advent of the DAW provided users a familiar template for audio interaction. With origins in the software revolution of the 1970s, its design elements persist in today’s interfaces that span from the film industry to scientific research. More recently, the rise of programming literacy and the expansion of audio research have evolved alongside the need and interest in low-level control. Libraries such as Librosa (Python), Web Audio API (JavaScript), and tuneR (R) have enthusiastic online userbases that connect across the internet, and the world. The broadening scope of creative coding bridges science and art to expand the worlds of the technical and the expressive. Apps and games built to facilitate music-making and sound exploration proliferate, and sonic arts has become well-established as a legitimate commercial field. In short, the spectrum of use cases in which audio is being transformed from numbers into something else is ever-expanding, and so, too, must the ways in which we interact with it. 

\section{A Brief History of Audio Visualization}
Modern audio analysis software has been continuously refined over the last century or so. Early hardware inventions that modeled sound signals were built using analog electronics to implement theoretical concepts from harmonic and spectral analysis. Ranging from exploratory to practical, these devices physically embodied the knowledge of sound as a medium of the times. They also carried with them necessary limitations and operational conventions that persisted in the shift from analog to digital. In today’s software, such assumptions are now often overlooked as analog origins were superseded by their digital descendants. DAW-like analysis software is at the heart of audio workflows that propel scientific inquiry, however, their embedded presets often assume specific use cases. These presets, since hidden, can easily be used to generate results using parameters intended for another domain. To better assess the contemporary landscape, we first review the historical origins of modern audio visualization tools. 

\subsection{The Origins of Sound Science}
Fourier’s seminal works on harmonic analysis (1807, 1822) [1,2] laid the mathematical foundations for audio signal processing, yet practical applications of these theories took time to crystallize. The earliest mechanical devices to record and play sound were the phonautograph in 1857 [3] and the phonograph (1877) [4], both of which were powered by hand. The phonautograph recorded sound waves by etching them on glass or paper [3]. The phonograph etched its sounds on tin foil, and could play back audio from the etching [4]. The invention of the telegraph (1837) [5,6] marked a transition as the first electric device to transmit sound encoded as signals, paving the way for the telephone (1876) [7], gramophone (1887) [8], loudspeaker (1925) [9], and sound spectrograph (1946) [10]. Friction and inertia of mechanical parts, short-circuits, and overheating are but some of the factors that impacted their smooth operation. These limitations were far from hidden: they were explicit, tactile, and fundamentally affected how users interacted with and interpreted sound. 

\subsection{Theoretical Foundations: Let’s Get Digital}
The development of the Fast-Fourier Transform (FFT) in 1965 [11] formed the backbone of signal processing algorithms as digital computing became ubiquitous through the rest of the century, and beyond. FFT-based methods impacted a wide variety of industries, for example, telecommunications (DSL modem [12], cell phones [13]), medicine (MRI [14], EEG [15]), and music (reverb [16], phase vocoder [17,18]). Along with music production, speech analysis, and sonar engineering, the impacts of applied Digital Signal Processing (DSP) radiated outwards, gradually becoming incorporated into the greater lexicon of digital audio analysis software, where they now live side-by-side as part of an unassuming digital toolkit. 

\subsection{How We Interact With Sound: Interface Design and the Rise of the DAW}
Arguably, the first DAW was the Soundstream Digital Editing System (1977), which operated on a minicomputer that ran custom software called the Digital Audio Processor (DAP) [19,20,21]. It was designed to edit master tapes, and featured hard disk recording, an interactive screen for waveform editing, and both analog and digital interfaces [20,21]. The Fairlight CMI (1979) was the first polyphonic synthesizer, well-known for its “Page R” sequencing environment that displayed rows of blocks that represented notes and audio [22,23]. Text-based DAWs, such as the Commodore 64 (1982) [24] and Keyboard Computer System (KCS) (1984) [25,26], supported multiple MIDI tracks using lists and drop-down menus. However, it was the Steinberg Pro-16 (1986), a software interface developed for the Atari, that had a visual layout closely resembling today’s DAW interface [27]. It looked like a physical hardware mixing console, complete with playback and routing controls, and horizontal arrangement views [28]. Computer processors at the time could not yet support multi-track recording or playback due to power and space limitations, so these early workstations were MIDI-only. Throughout the 1990s, the semiconductor industry enabled processor technology to become cheaper, faster, and smaller, enabling audio workstations to combine multiple features into the same device. Examples of early multifunctional software include Sound Tools (1989), with its limited audio recording [29,30]; Cubase (1992), with its MIDI and audio visible in the same interface [31]; and Virtual Studio Technology (VST) plugins (1996), which allowed digital effects to be applied to individual channels [32]. 

\subsection{How We Perceive Sound: Sensory, Perceptual, and Cognitive Considerations}
For most humans, sound is one of five core senses we experience throughout our lives. Our relationship with it changes as we age, and as we add information to our sensory network through lived experiences. A number of tools are used to visualize sound, some of which strive to depict spatialized relationships between its components, and others which employ layers of abstraction to expand its sphere of perceptible information. Oscilloscopes plot time-amplitude waveforms by reading the voltage from a transducer (microphone) to display pressure oscillations [33]. A spectrogram uses the Short-Time Fourier Transform (STFT) to sum windowed segments of a signal, trading temporal precision for frequency resolution: lower time-resolution allows the calculation of finely-grained frequency evolution, and vice-versa [34]. Mel-Frequency Cepstral Coefficients (MFCCs) represent spectral energy as a series of coefficients scaled exponentially to align with the human auditory system [35]. These representations are optimized for quantitative feature extraction, however, they can obscure more nuanced structures such as the timbre of a unique voice, the microtonality of an oud, or the rich polyphony heard while standing in the middle of a crowded train station.

\subsection{Dimensional Representation and Experimental Media}
One major challenge in data visualization is mapping high-dimensional features to visual variables in a way that intuitively makes sense when you look at it. Tools from statistics, such as scatter plots and time-series graphs, are precise and well-established, yet they require an input of low-dimensional data. Audio features, which are highly multidimensional (e.g. dozens of MFCCs, spectral and temporal centroids, entropy scores), require correspondingly advanced encodings. There are innovative efforts across many domains that strive to expand and explore the nature of data visualization, and to unify multidimensional and interactive visualizations with cognition. For example, topological data analysis (TDA) can reveal the underlying shape of a dataset [36,37], and has been used in describing the periodicity of flutes [38], music tagging and classification [39], and audio fingerprinting of MIDI music [40]. These shapes can then be fed into a convolutional neural network (CNN) as training data to teach it to detect patterns in audio features, which are output as activation maps [41]. 

As experimental graphics research continues to push the boundaries of technology, media domains such as virtual reality (VR), augmented reality (AR), mixed reality, and 360 video offer expanded formats for multisensory immersion. These technologies, often referred to as experiences, prioritize interactivity and can be found in spaces from VR gaming centers to live theatre and performance art. Societal applications include the use of haptics to enhance sensory awareness for blind or deaf people [42,43], VR for therapy and training [44], and 3D sculpture as a tool for design [45]. One important audio research application uses 3D time-frequency embeddings to visualize timbral similarity by projecting features into a spatial manifold, visualizing clusters of similar bird calls or phonetic units [46,47]. In a more exploratory vein, sonic labyrinths use interactive 3D structures to represent sound, where navigation corresponds to spectral exploration [48]. Across science and media, innovations in audio data visualization proliferate as technology facilitates the accessible transformation of multisensory information.

\section{Addressing Specific Knowledge Gaps}

\subsection{Hidden assumptions: software as a black-box}
The metaphor of the black-box comes from a fusion of aviation industry and WWII-era slang, when flight data recorders, along with other secret electronic devices, were housed in nonreflective black metal boxes [49]. While the first flight recorder used a thin beam of light to record metrics such as altitude and speed onto photographic paper, later versions engraved them onto metal foil [49]. The black-box metaphor has since become an analogy for the study of a closed system without prior knowledge of its inner workings, relying solely on knowledge of input, and observation of output, to evaluate its structure and evolution [50]. 

Comprising anywhere from hundreds to ten-thousands of lines of code and more, it becomes necessary to treat software as a black-box, or we would never get anything done. Since code is more often read than it is written [51], especially for free, libre, and open-source software (FLOSS), it is seen as a best practice to leave a clear, well-documented paper trail in the form of in-line notes, for posterity. Along with a (hopefully) clear set of instructions on how to use the software, these notes, known colloquially as documentation, are essential so that others who use it thereafter can follow the design and flow of logic, understand unimplemented features, or participate in future scaling efforts. Documentation facilitates both a deeper understanding of such tools, and the ability to change, edit, or repurpose software for permissible uses under the published license. Furthermore, for developers who may often work intensively in solitude, documentation serves as a form of communication and connectedness between people who may never meet each other in real life, adding an additional layer of meaning aside from utilitarian need. 

\subsection{Parameters, presets, and preconceived notions}
Transparency in software design facilitates access to customization that may liberate the user from the constraints of domain-specific applications. Knowledge of equations from fields like signal processing, population dynamics, or neuroscience can permit a user to trace the flow of logic through an ocean of code. Portability and translatability are also facilitated by such transparency, and at times it can be easy to replace one equation with another to achieve a new end goal. Code translations of such equations are often direct, if dense, mathematical translations through layers of abstraction known as standard software libraries (e.g. numpy, librosa, fftw). As with all equations that govern the empirical sciences, numerical parameters must be chosen to allow mathematical computation to occur. However, as meta-uses compound, the implicit reliance on presets or parameters can become buried, obscured, or forgotten. Therein runs a risk of making assumptions that may not be appropriate for a specific domain’s application. In the following section, we focus primarily on a comparison of FLOSS tools and their hardcoded assumptions (See Appendix A,B). 

\subsubsection{Presets and Defaults}
Praat was developed specifically to study the human voice, and has pre-emphasis filtering that boosts frequencies above 50 Hz. This alters the relationship between frequency content in the signal, and can be problematic for the study of animals that communicate using low-frequency information, such as whales, elephants, tigers, and rhinos [52-55]. It also limits the visual display of audio clips over a certain length. 
In scikit-maad, a 4th-order Butterworth (infinite-impulse response) filter is the preset for automated feature and region of interest (roi) selection. This filtering optimizes frequency precision with a flat passband and -24dB/octave rolloff, but limits temporal precision due to its phase-nonlinearity. Since different frequency components of a signal travel at different rates, this shifts the timing of low- and high-frequency information differently within the same acoustic event. The infinite filter response can also create acausal pre-event artifacts that interfere with the detection of onset transients. To mitigate this, maad defaults to the zero-phase filtfilt, but this choice is inappropriate when high temporal precision is needed. Examples include measuring intervals between syllables (such as echolocation clicks), sample-level accuracy for onset detection, or fine-scale waveform comparison. Using scipy.signal can allow for better control.
Librosa’s native sample rate is set to 22.05 kHz, and its STFT parameter defaults are set to a nfft value of 2048 and hop length of 512. Unless you know about this, you may be performing calculations with incorrect assumptions. 

\subsubsection{Workflow and Ergonomics}
More fully-featured software, such as Audacity, Sonic Visualiser, Avisoft (proprietary), and Raven (proprietary), represent a spectrum of graphical DAW-like tools that have developed specialized use cases in audio information domains. Their workflows are rooted in temporal manipulation, which is often (but not always) a stepping-stone in audio information science. For example, the purpose of cutting audio at annotation points is to then perform other calculations on that audio slice, i.e. feature extraction.
Horizontal vs. vertical layouts are tied to workflows from the audio recording industry. For scientific use cases, comparing many small files along horizontal timelines feels clunky when looking to broadly assess their similarities and differences. This is different from when we want to view the audio as a time sequence, where (horizontal) temporal continuity may be useful.
Interacting with all files (or annotated slices) at once can be labor-intensive, often requiring manual interaction with each one. There is not always a way to batch import many files vertically along independent channels. Files may be required to be loaded individually, or the batching of such files might be for a calculation or analysis that is hidden in the software’s algorithms.
If batch loading and viewing is indeed possible, interacting with all files simultaneously can require the manual labor of clicking each single track to turn such a feature on. Repetitive clicking with a mouse or trackpad is not physically ergonomic and can cause repetitive stress injuries. 
For effects batching, this is further exemplified. If a bandpass filter is required to eliminate some machine noise or a natural event such as an earthquake, it is far more efficient to apply this same effect to all files at the same time. Instead, one might have to manually click a checkbox, button, or VST device onto every channel – a task that quickly becomes tiresome or prohibitive for thousands of files. 

\subsubsection{Algorithmic Transparency and Limitations}
Audacity’s power spectrum calculation limits nfft value choices based on signal length; as such, the same nfft value can’t be chosen for all files in a batch if they are of non-uniform lengths. Also, spectral analysis can only be performed by clicking through a series of sub-menus, and can only be done on one sound clip at a time. The low-level libraries that supposedly allow for batch processing of files to do this task don’t actually work as described in the online documentation.
Audacity’s Fourier transform (pffft) relies on a translation of Fortran 77 code from FFTPACK that was written in 1985. These algorithms are very powerful, but may be difficult to integrate with modern software, and may not behave as expected, since they were designed to operate on hardware that had different limitations.
The number of different FFT algorithms that have been written and re-written for specific uses is at this point an unofficial meme in signal processing. This is evident across many different packages with amusing names such as “Pretty Fast Fast Fourier Transform” (pffft), “Keep It Simple Stupid Fast Fourier Transform” (kissfft), “Fastest Fourier Transform in the West” (fftw), and others. This can be overwhelming to keep up with when choosing algorithms.
Numerical computation always contains hidden assumptions that form a collection of presets, whether for parameter values, expected modes of user interaction, or conceptual approaches to sound. Indeed, tool choice is often made based on the baked-in assumptions that align most closely with a task at hand. This is neither inherently good nor bad, but a phenomenon of engaging in real-world problem-solving. 

\section{PROPOSED SOLUTIONS}

\subsection{Design Principles}
In the previous section, we outlined a technical wish-list based upon issues we have encountered in our use of audio analysis software. Informed in tandem with historical perspectives and conceptual extensions, we present a variety of solutions to the problem of tradeoffs due to the inherent uncertainty in information knowability. These go beyond solving technical issues into an evaluation of the landscape of contemporary cognition. We propose that giving users access to independence and agency facilitates an increased ability to form complex cognitive associations. (In a sense, this concept moves slightly outside of software into the domain of pedagogy, however, we strive to refine our focus toward the field of audio information visualization.) In the argument for this proposed solution, we identify three fundamental principles at the core of our design philosophy. 

Transparency – a clear-box approach, rather than a black-box approach, can empower the user to make their own appropriate choices for their intended use. This can involve presenting available options as visual cues at the point of interaction, rather than making decisions for the user or simply leaving all instructions in the documentation. It could also involve informing the user as to why certain design choices were made, and provide options for real-time reconfiguration. 

Flexibility – the ability to configure an environment that best aligns with an individual’s task requirements or work style can give a sense of agency over workflows. Sometimes, it is especially useful to have multiple perspectives when trying to understand a complex situation. The difficulty of working with time-series data is no exception; the ability to switch seamlessly between analogous options, and even to compare them side-by-side or embedded upon each other, can be very informative. Adaptable design principles make tools easier to use across a wide variety of scenarios, and may encourage users to stick with one familiar tool, rather than switching frequently between divergent workflows. 

Robustness – tools should handle a wide variety of contexts, and be as agnostic as possible to types of data input. This could mean that a tool is designed to process input data in many ways, like a hammer, or to receive and combine many types of data in a synthetic configuration, like a multi-tool. Consider software that is designed to receive uniform lengths of audio from the same source. A next step might be to map extracted features and combine them with environmental variables, such as weather and temperature, or with metrics taken across the set of input data, such as mean amplitude or spectral centroid. The raw data itself already has a certain uniformity, so the parameter space of this tool would then be highly synthetic, since it would be constructed out of higher-order relationships between abstract variables. Alternatively, if a tool’s input sounds have high heterogeneity, like clips of drastically different lengths or sounds from different species, efforts to generate a base parameter space might first focus on defining broader sets of classical metrics, such as duration, amplitude, entropy, or various other statistics prior to abstract transformation. The key difference between these two scenarios lies in the input data. Each requires a different number and types of steps to transform data to the same point of abstraction. Robust tools should be configurable for either case. 

\subsection{Conceptual Design Principles}

The theoretical benefits of incorporating an updated set of modern design principles into audio visualization workflows have far-reaching implications outside of simply being less annoyed while performing daily tasks. Studies across cognitive psychology and design theory show that increased perceptual connections can enhance pattern recognition [56-58]. The following examples demonstrate how spatial and temporal representations of information impact mental processes such as comprehension, memory, and learning. 

\subsubsection{Cognitive Load Theory} Split-attention effects show that having to combine information from multiple, individual, spatially-separated sources inhibits learning [56]. These effects are also found in scenarios where information is presented simultaneously, but in different formats [56]. This implies, conversely, that if information is visibly close together, and/or presented simultaneously but in the same format, learning will be easier. In audio software, we can draw an analogy to split-screen views that show waveforms, spectrograms, and power spectral density on separate screens. Users are required to constantly switch back and forth between views, trying to remember what they previously saw on the last screen as they translate information from one format to another. (This is an actual, real problem in Audacity; see section IV-b.) Such display issues limit a user’s mental availability to make intuitive inferences, since one must search for and map visual elements back to each other while holding prior information in working memory. The demands on cognitive load also increase when information is presented sequentially [57,58], rather than in staggered or simultaneous formats. Furthermore, information complexity is modulated not just by the total number of elements, but also by their interactions [57]. Simultaneous information streams require greater load on working memory [57,58]; therefore, the more interconnected a group of elements is, the more complex the information they represent. From this, we can conclude that sequential formats are not ideal for processing complex interconnected information. Outside of cognitive psychology, inefficiency in linear and sequential information processing has been shown in the communications [59], computing [60], and energy [61] industries. Since humans are the architects of these systems, the phenomenon that preferences a non-simultaneity of information processing could even be a function of human cognition, but that is outside of the scope of this paper to explore. 

\subsubsection{Visual Design} The effects of visual elements on perception have been explored systematically through a variety of principles that govern design theory. The visual variables framework describe position and size as the principal factors that express quantitative differences [62]. Color, as a variable, is broken into the values of hue, which describes the qualitative difference of category, and value, which describes the quantitative difference of order [62]. Together with shape, orientation, and texture, these visual variables describe a hierarchy of information with levels that are either associative or dissociative [62]. This means that visual characteristics can be used to deconstruct the emergent patterns that inform meaningful group characteristics. That is to say, when objects are perceived as being part of a group, visual variables provide a basis for distinction. To extend these thoughts to audio software and visualization, we can thereby conclude that the ability to identify patterns in abstract representations, such as those used for audio visualization, can be facilitated by making visual design choices that correctly map visual elements to meaningful features. This is consistent with existing approaches for dimensionality reduction in modern data visualization.

\subsection{Jellyfish Dynamite}
\begin{figure}[htbp]
    \centering
    \begin{minipage}[b]{0.45\textwidth}
        \centering
        \includegraphics[height=6cm]{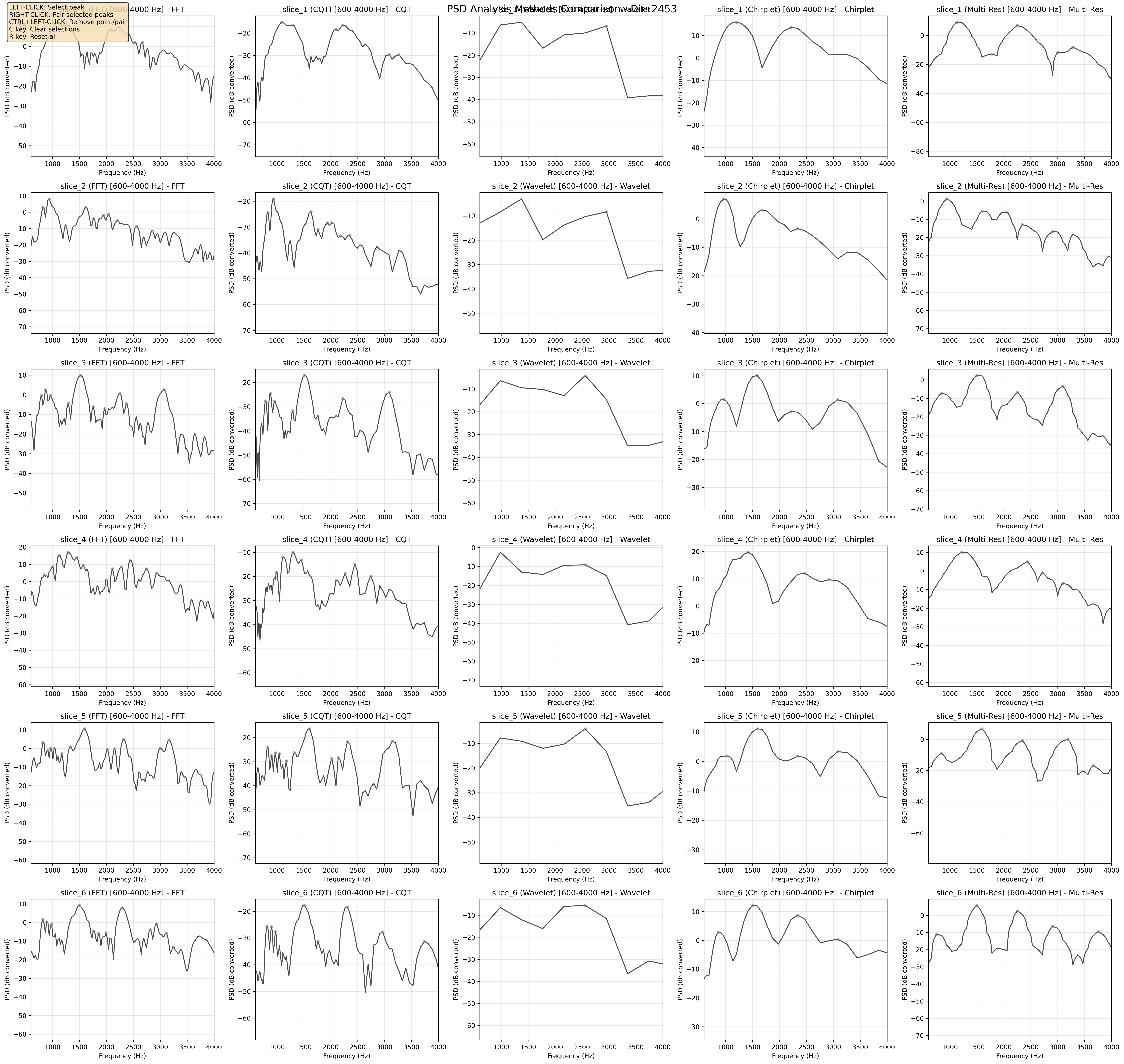}
        \captionof{figure}{Comparison of power spectral density transformations using FFT, CQT, wavelet, chirplet, and multi-resolution methods, displayed horizontally, for a time-sequence of audio syllables, displayed vertically.}
        \label{fig:psd_multiplot}
    \end{minipage}
    \hfill
    \begin{minipage}[b]{0.52\textwidth}
        \centering
        \includegraphics[height=6cm]{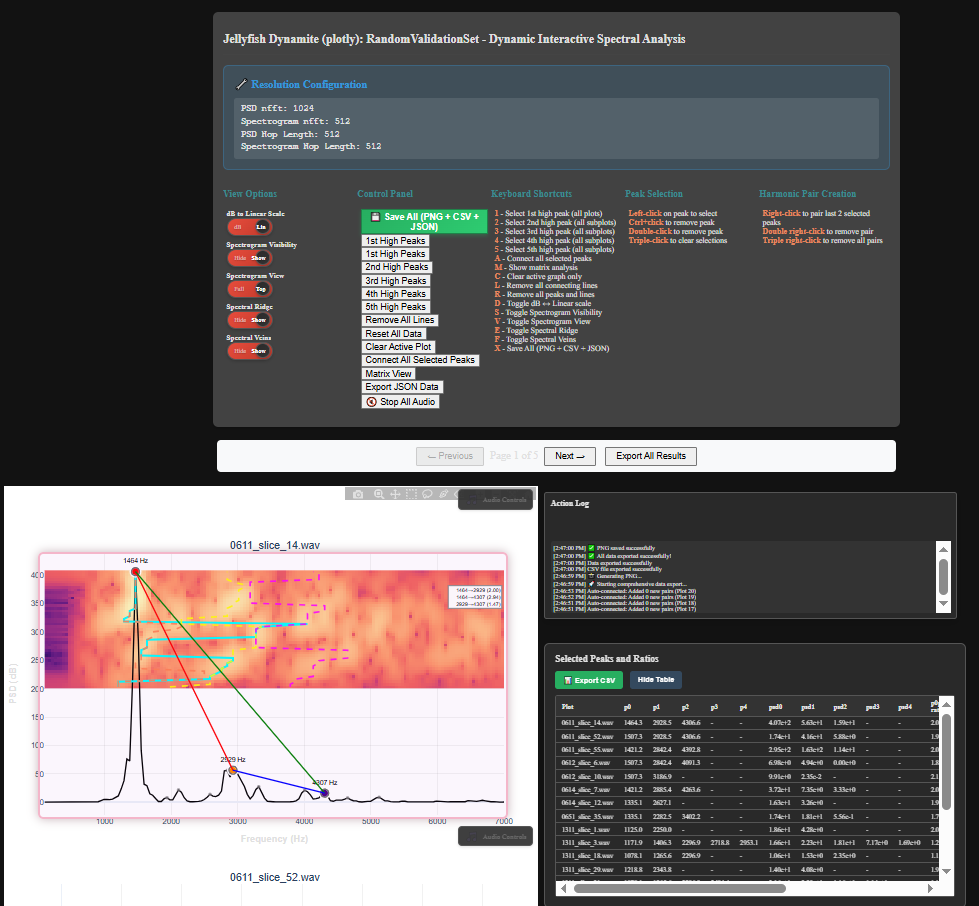}
        \captionof{figure}{Jellyfish Dynamite Interface. Plot shows an audio power spectrum with spectrogram overlays, peak connections, and energy tracking lines. Interface controls contain buttons, switches, and instructions for use. Data tables contain ready-to-export peak selections.}
        \label{fig:jellyfish_interface}
    \end{minipage}
\end{figure}

\begin{figure}[htbp]
    \centering
    \includegraphics[width=0.8\textwidth]{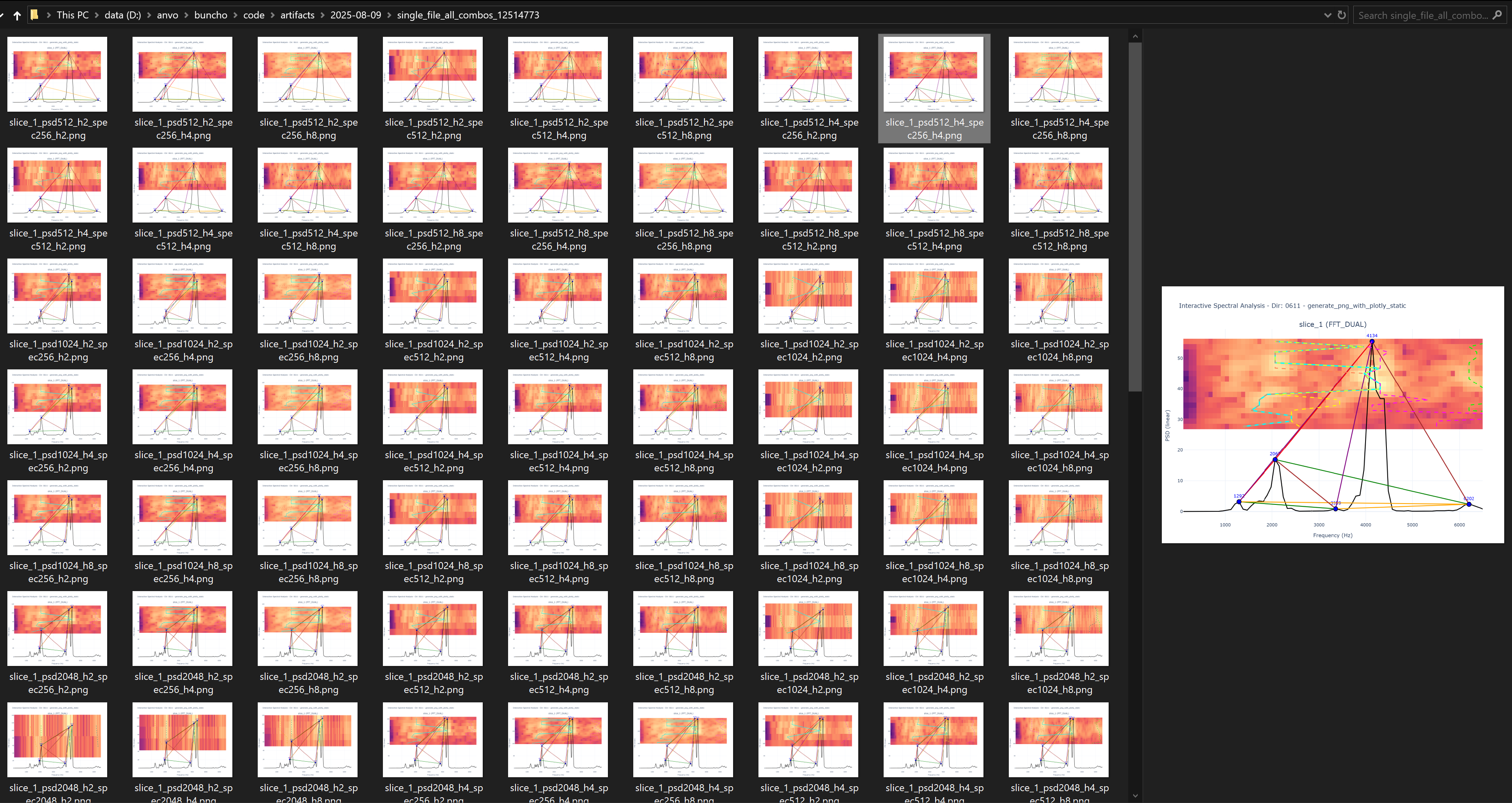}
    \caption{(Left) Full comparison of dual-scale spectrogram selections visualizes every possible combination for nfft values of 512, 1024, 2048 and hop length of 2, 4, 8. (Right) Fully-connected peak plot showing PSD, spectrogram, and energy ridges for a single audio file from Jellyfish Dynamite's interactive interface.}
    \label{fig:dual_scale_spectrograms}
\end{figure}

In an effort to address the issues we have discovered in our research, we designed a software solution to the problems outlined in this paper. Jellyfish Dynamite is an extensible, interactive Messerolle for audio data information visualization. Written in Python, its preprocessing stage segments audio at annotation points and structures syllables into a bird pair dictionary, retaining metadata as a standalone dataframe. The backend processes audio using custom algorithms to compute high-level features, and implements several spectral transformation algorithms (\texttt{FFT\_DUAL}, \texttt{CQT}, \texttt{WAVE}, \texttt{CHIRPLET}, \texttt{MULTI\_RES}) (Fig.~\ref{fig:psd_multiplot}) that each return unique frequency bin and corresponding PSD (power spectral density) magnitude values. The frontend is an interactive interface (Fig.~\ref{fig:jellyfish_interface}) that supports multiple views (Fig.~\ref{fig:dual_scale_spectrograms}) and computes transformations using keycommands and mouse-click combinations. Peak frequencies can be automatically computed with up to four peaks initially auto-filled on plots. Users can then interact with the peaks to deselect them, as well as to select additional peaks. There are a number of buttons that change the views and scales of the visual display. Selected peaks are added to a computation table in real-time, and harmonic ratios are displayed visually on the plots. Data is exported as any mutually-inclusive combination of csv, json, or png files. The overall architecture of the interface uses a MVC (model-view-controller) structure, where the model is a data array with spectral values, the view is a set of plots and controls, and the controller is a set of event handlers that implements precisely timed interactions between the user and the data model. (See Appendix C and \url{https://github.com/laelume/jellyfish_dynamite} for reference.)

\section{DISCUSSION}
\subsection{Practical considerations}
Through the lenses of cognitive and visual design theory, we show that associations between visual elements and the human psyche are intrinsically linked through the perceptual continuum that is bodied sensory experience. The inner workings of human cognition and psychology fundamentally demand an interactive format to give context to complex information. We can therefore project that for audio information visualization design, users may benefit from access to tools and workflows that allow for a perceptually diverse engagement with sound. This could include nonlinear workflows, reorienting information along different axes, using new metrics to scale information, or interchanging relationships between variables. The incorporation of contemporary design principles into audio analysis tools and workflows can expand the boundaries of both technical analysis and creative sound exploration. Practically, it takes time to implement new tools. Novel visualizations may require a shift in representational paradigms: new information is not always readily accepted. To be fully adopted, users must first overcome cognitive dissonance and resistance to change [63,64], followed by the learning curve that is associated with performing any new task. As familiarity and then mastery is attained, these tools can become streamlined into existing workflows. We may even struggle to remember what life was like before we had access to them; such is the curse of convenience. However, increased technical literacy begets the benefits of speed, efficiency, and creative flexibility. 

\subsection{Future impact, intended audience: who benefits?}
There are endless ways to explore the theoretical effects of applied design philosophy, but what about their impact? When a new tool or technique is deployed, who will actually use it? Who will it benefit? Where and how will it be used? Especially now, in the age of Big Data, there is an accelerated need to include non-domain experts and citizen science participants in the validation and annotation of data. Tools designed specifically with interaction and visualization in mind can make it more accessible for people to interact with data in ways that are relatable, intuitive, and familiar. The tactile experiences of everyday digital tools, such as apps and games, can be modeled and expanded upon to create user experiences that feel familiar while not being too distracting. Such tools can also give people a sense of agency over what they’re doing – they may reveal the ‘secret elements’ that are often reserved for specialists, increasing transparency, building institutional trust, and generating a sense of community investment. Furthermore, tools that are fun and interesting to use generate conversations outside of their initial use/community. When everyday people get excited enough about wild bird audio annotation apps to discuss them at coffee shops or networking events, for example, this can be viewed as a sign of success that such a tool is connected with social values. Thus, there are diverse practical reasons in favor of increasing the accessibility of audio analysis and exploration to both technical and non-technical audiences. The following are some examples of benefits to specific groups: 

\subsubsection{Professional and Scientific Users}
People who already use data visualization tools regularly for their jobs, such as scientists, data scientists and analysts will certainly benefit from increased efficiency and intuition, allowing them to see audio information in new ways. Specialized task automation, efficient 3D or time-evolution displays, and the ability to visually overlay features of interest in new ways are some hypothetical workflows that could be beneficial.
AI users in particular, who may not be used to working with noisy real-world data, or who may work with many different types of data, require assistance in understanding the nuances of datasets when they are not familiar with the subject matter. In the rising proliferation of AI outside of experimental and research domains, the number of people working with audio data will increase dramatically, as will the use of AI as an everyday tool in its own right. Such human individuals (and, more dangerously, their AI counterparts) can make incorrect assumptions about properties or characteristics of sound if they are not informed in a way that is fast, efficient, and intuitive. This also factors into the field of ethics, since the dangers of making assumptions can proliferate quickly in cases where a small effect may spiral out of control over a massive dataset like those seen in Big Data, or may propagate into models through training, or affect other datasets through extracted metadata.
\subsubsection{Citizen Science and Public Engagement}
Citizen scientists who participate in valuable tasks such as data annotation and validation, species identification, symptom reporting, noise pollution assessment, can have a way to easily annotate in real-time that may allow them to feel included as an essential part of a team, gives them more knowledge about the science and behind the scenes, which could encourage them to become more excited and involved from a scientific standpoint. This is triply beneficial because science education is essential as people need to work together to address many urgent problems in fields such as conservation, medicine, and society.
Accessibility by including things that are interesting or fun to look at, listen to, and interact with, especially for non-experts, can provide entertainment as well as social values. The possibility of gamification can also increase audience reach, and can be used to collect feedback about what does and doesn’t work, as well as who tends to use the tools and how, which are valuable insights for any tool designer.
\subsubsection{Collaboration and Shared Workflows}
We can imagine a use where, for a large dataset that needs annotation, it could be broken up into smaller pieces and distributed among a group of people to lessen the workload. Then, it becomes essential for all users to be sure they are referring to the same phenomena, and the same features, across the same interface.
Audio visualization tools can also act as intermediary steps between the many people involved along the way in the process of scientific and artistic inquiry. It places control in the hands of the user, and reconfigures the hierarchy that limits niche knowledge to be held solely by domain experts. Increased agency can build a sense of community, and strengthens the ties that people feel to their work or special interest.
Rapid advancements in audio data visualization are expected as the Age of Information spirals outwards. We hope that considering the implications and impacts of new tools on their audiences, the case for incorporating a broader set of user-centric design principles may be compelling. 

\section{CONCLUSIONS}
Sound as a phenomenon presents infinite possibilities for interpretation. Its analysis employs a wide range of tools, each carrying conventions that shape the ensuing frameworks of its representation. We assert that visualization can be framed as a set of analysis techniques that has become indispensable to the study of audio data. Since the human experience of sound perception is inherently multidimensional, mapping audio features into visual parameter spaces should reflect this complexity. Classic visualization tools might carry presets or conventions that interfere invisibly with information processing by using assumptions transferred from different applied contexts. To address these concerns, we have proposed the introduction of new or updated software tools that use transparency, flexibility, and robustness to better align with the domain-specific needs of modern audio analysts. Like a mountain range when viewed from a different angle, new perspectives can offer new insights. It is our hope that in the adoption of such strategies, we may facilitate an environment that allows us to see beyond sound.

\bibliographystyle{plainnat}
% End footnotesize

\appendix

\section{Domain Assumptions of Audio Software}
\label{app:domain-assumptions}

\begin{table}[htbp]
\centering
\footnotesize
\begin{tabular}{p{2.5cm}p{3cm}p{2cm}p{3cm}p{2.5cm}}
\hline
\textbf{Library/Software} & \textbf{Parameter/Setting} & \textbf{Specific Values} & \textbf{Domain Assumption} & \textbf{Use Case} \\
\hline
scikit-maad & Bandpass Filter & 1-8 kHz & Species vocalize in this range; noise exists outside it & Species Detection \\
scikit-maad & FFT Size & 512/1024 samples & ~23ms window balances frequency/time resolution & Spectrogram Analysis \\
scikit-maad & ROI detection & Variable thresholds & Biological sounds are contiguous energy blobs in specific bands & Species Detection \\
Librosa & Default sample rate & 22.05 kHz & Human-audible focus, STFT-centric worldview & Audio Processing \\
Librosa & STFT parameters & n\_fft=2048, hop\_length=512 & Standard frame-based analysis with Mel scale relevance & Feature Extraction \\
Praat & Pitch settings & 75-500 Hz & Source-filter model with human speech frequency ranges & Speech Analysis \\
Audacity & Default sample rate & 22.05/44.1 kHz & Human-audible and most animal sounds below ~10 kHz & Audio Processing \\
Raven Pro & FFT Size & 512/1024 samples & Standard trade-off for most animal calls & Spectrogram Analysis \\
\hline
\end{tabular}
\caption{Selected domain assumptions embedded in audio analysis software (non-exhaustive list).}
\label{tab:domain-assumptions}
\end{table}

\section{Extended List of Audio Software}
{\small  % Start smaller font

\label{app:software-list}

\subsection*{Python}
\texttt{Librosa}, \texttt{PyAudio}, \texttt{TorchAudio} (PyTorch), \texttt{fftw}, \texttt{affft}, \texttt{scikit-maad}, \texttt{pywt}, \texttt{pffft}

\subsection*{C/C++}
\texttt{Essentia}, \texttt{JUCE}, \texttt{Maximilian}

\subsection*{JavaScript/Web}
\texttt{Web Audio API}, \texttt{Tone.js}, \texttt{wavesurfer.js}

\subsection*{R}
\texttt{tuneR}, \texttt{soundgen}, \texttt{seewave}, \texttt{warbleR}, \texttt{monitoR}

\subsection*{Bioacoustics-specific}
\texttt{Praat}, \texttt{Parselmouth}

\subsection*{Sound Art}
\texttt{SuperCollider}, \texttt{PureData}, \texttt{Faust}, \texttt{ChucK}

\subsection*{DAW-like}
\texttt{Audacity}, \texttt{Sonic Visualiser}, \texttt{Raven}, \texttt{Ableton}, \texttt{Reaper}, \texttt{GarageBand}, \texttt{Logic}, \texttt{Pro Tools}

\subsection*{Experimental}
\texttt{Max/MSP}, \texttt{ORCA}, \texttt{FoxDot}, \texttt{Tidal}, \texttt{Sonic Pi}

\section{Jellyfish Dynamite Overview}
\label{app:jellyfish}

\subsection{Backend – Audio Analysis}

\subsubsection{Data Preparation \& Transformation}
\begin{itemize}[leftmargin=*]
\item Selects audio files. Filters .wav files from a directory structure based on user-defined indices, ranges, or filename patterns.
\item Applies multiple spectral transformations. Processes each audio signal through independent algorithms to generate comparative frequency-domain representations.
\item Executes \texttt{FFT\_DUAL} transformation. Computes a high-resolution PSD for frequency analysis and a lower-resolution spectrogram for time-frequency visualization.
\item Computes Constant-Q Transform (CQT). Generates a logarithmic frequency scale representation.
\item Performs Wavelet Packet Decomposition. Utilizes wavelets (sym8, db8) for multi-resolution time-frequency analysis.
\item Calculates Stationary Wavelet Transform (SWT). Executes a shift-invariant wavelet transform for enhanced feature detection.
\item Runs Chirplet Transform. Correlates the signal with frequency-modulated chirps to identify non-stationary components.
\item Constructs Multi-Resolution PSD. Stitches together results from FFTs of different window sizes into a continuous full-spectrum estimate.
\item Validates output data. Checks for and corrects NaN, Inf, and zero values to ensure mathematical integrity of all transformed data.
\end{itemize}

\subsubsection{Feature Extraction \& Detection}
\begin{itemize}[leftmargin=*]
\item Detects spectral peaks. Identifies local maxima in the PSD using adaptive thresholding based on height percentile and prominence.
\item Calculates peak properties. Measures the width, prominence, and power of each detected peak.
\item Finds maximum energy ridge. Analyzes the spectrogram to identify the dominant frequency trajectory over time.
\item Identifies spectral veins. Detects multiple persistent energy bands within the spectrogram by tracking local maxima across time-frequency windows.
\end{itemize}

\subsubsection{Interactive Analysis \& Graph Construction}
\begin{itemize}[leftmargin=*]
\item Presents multi-plot interface. Renders a grid comparing different files and methods simultaneously.
\item Maintains dual-scale system. Stores and manages both linear and decibel (dB) representations of the PSD data for instantaneous scale toggling.
\item Handles user input events. Processes mouse clicks (select, deselect, remove) and keyboard commands for analytical operations.
\item Constructs graph networks. Builds \texttt{networkx.Graph} objects where nodes represent frequencies and edges represent harmonic relationships, annotated with frequency ratios.
\item Performs automated peak selection. Ranks detected peaks by power and automatically selects the top N (e.g., 1-5) peaks across all subplots.
\item Calculates harmonic ratios. Computes and displays the ratio between any two user-selected or auto-selected frequencies.
\end{itemize}

\subsubsection{Output \& Validation}
\begin{itemize}[leftmargin=*]
\item Exports graphical results. Saves the complete interactive figure as a high-resolution PNG.
\item Serializes analytical data. Exports all selected peaks, pairs, frequency ratios, and graph data to structured JSON files.
\item Generates machine-readable tables. Outputs peak and ratio data into CSV format for statistical analysis.
\item Produces interactive HTML reports. Creates standalone web pages with Plotly visualizations that retain interactive functionality.
\item Executes parameter optimization. Performs grid searches using \texttt{n\_fft}, \texttt{hop\_length} to empirically determine optimal processing settings for a given signal type.
\item Creates validation datasets. Implements statistical sampling to select random file subsets for method validation and quality control.
\end{itemize}

\subsection{Frontend – Interactive Interface}

\subsubsection{Data Handling \& Initialization}
\begin{itemize}[leftmargin=*]
\item Loads pre-computed data. Injects serialized Plotly figure data and configuration parameters from the backend Jinja2 template.
\item Parses spectral arrays. Extracts frequency bins, power spectral density (PSD) values, and peak locations for each subplot.
\item Initializes interaction state. Creates data structures to track user selections, frequency pairs, and graph connections.
\item Sets initial visualization parameters. Configures scale (linear or dB), spectrogram visibility, and spectral feature overlays based on default settings.
\end{itemize}

\subsubsection{Visualization \& Rendering}
\begin{itemize}[leftmargin=*]
\item Generates subplot grid. Creates a fixed layout of individual plots arranged in rows and columns.
\item Draws PSD traces. Plots the main power spectral density curve for each audio file and analysis method.
\item Renders detected peaks. Marks initial peak locations with gray circular markers.
\item Displays spectrogram overlays. Draws time-frequency representations as semi-transparent heatmaps behind the PSD traces.
\item Calculates and plots spectral ridges. Computes and displays the maximum energy trajectory across time for each spectrogram.
\item Identifies and draws spectral veins. Detects and renders multiple persistent energy bands as dashed lines.
\end{itemize}

\subsubsection{Interaction Management}
\begin{itemize}[leftmargin=*]
\item Processes mouse events. Handles left-clicks (selection), right-clicks (pairing), and double-clicks (removal) on all plot elements.
\item Maps screen coordinates to data values. Converts pixel positions to corresponding frequency and power values for accurate selection.
\item Finds nearest peaks. Calculates distance between click position and all detected peaks to determine user selection target.
\item Tracks selection order. Records the sequence of user selections for color assignment and visual distinction.
\item Manages paired frequencies. Creates and stores relationships between selected frequencies, including calculated ratios.
\item Updates visual elements. Dynamically adds, removes, or modifies markers, connecting lines, and vertical indicators based on user actions.
\item Toggles display scales. Switches all plots between linear and decibel representations without recomputing underlying data.
\item Controls feature visibility. Shows or hides spectrograms, spectral ridges, and veins based on user toggle commands.
\end{itemize}

\subsubsection{Audio Integration}
\begin{itemize}[leftmargin=*]
\item Initializes audio context. Prepares Web Audio API components for sound synthesis and playback.
\item Generates sine waves. Creates pure tones at specified frequencies corresponding to selected peaks.
\item Loads original audio files. Fetches and buffers source audio for direct playback.
\item Applies time-stretching. Alters playback rate of original audio while maintaining pitch.
\item Controls audio parameters. Adjusts gain, looping, and playback state in real-time.
\end{itemize}

\subsubsection{Output \& Export}
\begin{itemize}[leftmargin=*]
\item Populates data tables. Dynamically updates HTML tables with selected frequencies, power values, and calculated ratios.
\item Formats data for export. Converts internal data structures to CSV and JSON formats for download.
\item Generates static images. Uses Plotly's image export functionality to create PNG files of the current visualization state.
\item Saves application state. Preserves user selections and analysis state in browser-local storage for session continuity.
\end{itemize}
} % End small font

\end{document}